\documentclass{iaus}
\usepackage{graphicx}

\title[Infrared Imaging of PNe] 
{Infrared Imaging of Planetary Nebulae \\ from the Ground Up}

\author[Hora]   
{Joseph L. Hora%
  }

\affiliation{Harvard-Smithsonian Center for Astrophysics,
60 Garden St., MS-65, Cambridge, MA 02138-1516, USA \break email: jhora@cfa.harvard.edu\\[\affilskip]
}

\pubyear{2006}
\volume{234}  
\pagerange{1--8}
\date{2006}
\jname{Planetary Nebulae in Our Galaxy and Beyond}
\editors{M. J. Barlow \& R. H. M\'endez, eds.}
\begin{document}

\maketitle

\begin{abstract}
New ground-based telescopes and instruments, the return of the NICMOS instrument on the 
\textit{Hubble Space Telescope} (HST),
and the recent launch of the \textit{Spitzer Space Telescope}
have provided new tools that are being utilized in the study of planetary nebulae.  Multiwavelength, high spatial resolution ground-based
and HST imaging have been used to probe the inner regions of young PNe to determine their structure and evaluate formation mechanisms.  \textit{Spitzer}/IRAC and 
MIPS have been used to image more evolved PNe to determine the spatial distribution of molecular hydrogen, ionized gas, and dust in the nebulae and halos.

\keywords{Spitzer Space Telescope, infrared: ISM, ISM: molecules,
planetary nebulae: general, techniques: high angular resolution}
\end{abstract}

\firstsection 

\section{Introduction}

Infrared imaging has been an important observational technique for 
planetary nebulae (PNe) for several reasons.  First, the ejected matter
from the star often obscures the inner regions of the nebula.  In the 
infrared, one can penetrate the surrounding dust and gas and see into the 
regions near the central star, which may be key in understanding the 
formation mechanisms that lead to the PN structure.  Second, there are 
many infrared lines that are important tracers and diagnostics of shocks 
and photodissociation regions (PDRs).  These include the bright H$_2$ line
at 2.12 $\mu$m and many other lines from 2 -- 28 $\mu$m, the 
[Fe II] 1.64 $\mu$m line, and several other bright diagnostic lines such
as 4.05 $\mu$m Br$\alpha$, 12.81 $\mu$m [Ne II], and other lines.  Finally,
there are components of the nebula that can only be observed in the infrared.
These include the PAH emission features at 3.3, 6.2, 7.7, 8.6, and 11.2 $\mu$m,
and thermal continuum emission from hot, warm, or cool dust.

Since the last PNe symposium, observers have made use of improved facilities
and instruments to probe PNe in the infrared.  There are now several telescopes 
in the 8-10 m class in both hemispheres that provide a significant increase in
sensitivity and resolution over what was previously available.  The capability
of the instruments on these telescopes has also been greatly enhanced 
with large-format IR imagers, many with adaptive optics (AO) that can deliver nearly 
diffraction-limited image quality at longer wavelengths.  These systems utilize
narrow-band filters to isolate spectral features and nearby continuum wavelengths
to map the emission components of the nebula.  

Previous and current space missions have also provided new resources for 
infrared imaging.  Surveys such as the MSX survey of the Galactic Plane, and the 
all-sky 2MASS survey are important datasets that are being utilized to find
or study PNe in the infrared.  The HST/NICMOS instrument has been
used to study the compact structure of young PNe in the near-IR.  Finally,
the \textit{Spitzer Space Telescope} (\cite[Werner et al. 2004]{werner04}) was launched in August 2003 with 
three instruments that have imaging capability from 3 - 200 $\mu$m.

In this review I will summarize some representative published results since the last 
IAU Symposium, and some of the Spitzer GTO observations of PNe.  Although I
stress here the infrared imaging aspects of recent results, the IR emission is only 
one important part of the full picture of these PNe.  Many investigators employ a
multi-wavelength approach using images and spectra
that greatly enhances the understanding of the PNe's 
structure. 

\section{Ground-based Near-Infrared (1-5 $\mu$m) Imaging}

Early near-IR array camera imaging surveys 
(e.g., \cite[Latter et al. 1995]{latter95};
\cite[Kastner et al. 1996]{kastner96})
focused on narrowband imaging of various well-studied PNe, exploring the
relationship between H$_2$ emission and morphological type, and examining differences
between the near-IR and optical structure. 
Much of the recent ground-based near-IR 
imaging has concentrated on high angular resolution
narrowband observations of young PNe.  The goal has been to probe the compact
structures near the core and in the lobes of the PNe to gain an understanding
of the mechanisms that are creating the bipolar or asymmetric morphology.  

One example is \cite{volk04} who used the Gemini telescope with
the QUIRC instrument and Hokupa'a AO system
to perform high resolution (0.15'') 2.12 $\mu$m 
and narrowband continuum imaging of three young PNe, IRAS
19306+1407, IRAS 20028+3910, and IRAS 22023+5249.  They found the H$_2$ emission
primarily in the lobes or in an equatorial torus around the star.  They 
demonstrated that with ground-based near-IR imaging one can detect 
and resolve structures in objects that are in the very early 
stages of PN formation.

\cite{sahai05} applied similar techniques to another young PN, IRAS 16342-3814.
They used the Keck II telescope with NIRC2 and natural guide star AO system
to obtain $H, K_p, L_p,$ and $M_s$ images that show 
limb-brightened bipolar lobes with a ``corkscrew" pattern inscribed in the 
walls of the lobes.  They conclude that this pattern is the result of a 
well-collimated jet with a diameter $\leqq$ 100 AU and a precession period 
$\leqq$ 50yr. 

Ground-based near-IR imaging has also been used in conjunction with HST 
(and longer wavelength) imaging
to probe regions that are heavily extincted near the core and to 
trace the distribution of PAH molecules and hot dust emission.  
\cite{matsu05a} used the ISAAC instrument on the VLT to image the
bipolar PN NGC 6302 in the 3.28 $\mu$m PAH and Br$\alpha$ lines, as well
as continuum wavelengths.  The PAH feature emission is found to be similar to
the ionized emission, although in the PAH image
there is an additional local maximum in a ``spur'' of
emission
and a shell extending away from the core.
They also used the JCMT to obtain a 450 $\mu$m
continuum image which shows an extended massive dust disk around the central
core.  From these data, they were able to estimate the mass of the dust and the 
disk, examine the complicated structure in the core, and detect a point source 
that may be related to the central star.  \cite{matsu05b}
performed 
a similar study of the bipolar PN NGC 6537.

The observations by \cite{smith03} of Menzel 3 (Mz 3) show the power of 
combining narrowband images with optical and near-IR spectra.  Mz 3 is a 
bipolar nebula that is similar in many respects to M 2-9, except that Mz 3
is a bipolar with no detected H$_2$ emission.
With the spectra, Smith used the CLOUDY program to model
the central source temperature and determine abundances of the ejecta.
The images in particular lines
indicate how conditions vary across the nebula.  

When high spectral resolution
imaging can be performed, a wealth of information can be obtained on the
morphology and kinematics of the nebula.  This has been demonstrated for 
example in the PN NGC 7027 and AFGL 618 (\cite[Cox et al. 2002]{cox02}; 
\cite[Cox et al. 2003]{cox03}).  They used 
the BEAR instrument, an imaging Fourier Transform Spectrometer 
operated at the CFHT that provided a spectral
resolution of $\sim$9 km\ s$^{-1}$.  These data allowed Cox et al. to 
examine the kinematics of the ionized gas and molecular components, and
to probe the interactions in the outflows.

\section{Ground-based Mid-Infrared (8-25 $\mu$m) Imaging}

The recent availability of facility mid-IR instruments on large telescopes
has enhanced the ability to probe even deeper into obscured regions, and 
to detect emission from warm dust and molecules.  The objects which have
been most often observed are young PNe which are bright in the mid-IR and their
central regions are obscured by dust in the optical.  The first instruments 
based on sensitive mid-IR array detectors were developed in the late 1980's and
were used on the large IR-optimized telescopes of that time to study this 
class of objects
(e.g., \cite[Hora et al. 1990]{hora90}; \cite[Meixner et al. 1993]{meixner93};
\cite[Deutsch et al. 1993]{deutsch93}; \cite[Persi et al. 1994]{persi94};
\cite{kompe97}; \cite[Meixner et al. 1999]{meixner99}).
Some mid-IR imaging has continued to be
performed on 3-4 m telescopes. For example, \cite{kemper02} used TIMMI2
on the ESO 3.6m to image NGC 6302, and \cite{matsuura04} imaged the post-AGB
star IRAS 16279-4757. \cite{smith05} 
used MIRLIN on the IRTF to image M2-9, Mz 3, and He 2-104.  

Recently, mid-IR instruments on Keck, the VLT, and 
Gemini telescopes have been exploited to achieve higher sensitivity and resolution.
\cite{muthu06} used the Gemini South telescope with the T-ReCS instrument to
image Hen 3-401, a bipolar nebula viewed nearly edge-on, with lobes separated
by a dust lane.  With continuum images at 10.4 and 18.3 $\mu$m, they 
constructed a color temperature map of the emission in the lobes.  They obtained 
a narrowband image in the 11.3 $\mu$m PAH feature and determined that
the PAH emission is organized in concentric sets of arcs in the 
lobes, which suggests several ejection epochs.  In another investigation of mid-IR
spectral features, \cite{kwok02} used the Gemini North telescope 
and OSCIR to examine two 
proto-PN that possess the 21 $\mu$m emission
feature.  They found a similar morphology 
in the 11.3 and 21 $\mu$m emission, suggesting that the carriers of both features
originated in the material ejected during the AGB phase.

In another example of a multi-wavelength approach, \cite{lagadec06} used
archive HST optical data and near- and mid-IR data from 3 -- 13 $\mu$m to examine
the core of the PN Hen 2-113.   This PN had been imaged previously by \cite{sahai00}
who found the morphology to be bipolar, with two ring-like structures around
the central star location.  Lagadec et al. obtained their images with several
instruments (NACO, MIDI, ISAAC, and TIMMI) at ESO, and analyzed them in conjunction
with the HST data.  They determined that the PN has a ``diabolo'' structure,
with the previously observed rings being the opening of two cone-like structures defining the 
bipolar outflow, and a dark lane between them.  There are also many other structures
detected, such as bright regions and spots, and filaments and holes in other parts
of the lobes.  The distribution of the PAH emission leads them to conclude that
the PAHs formed more recently than the continuum-emitting dust.

\section{HST/NICMOS Imaging}
The installation of the NICMOS instrument on HST in 1997 
provided the capability for high
sensitivity, diffraction-limited
1-2.5 $\mu$m imaging with many narrow and broad filters, and was used for several
studies of PNe (e.g., \cite[Sahai et al. 1998]{sahai98} (AFGL 2688); 
\cite[Latter et al. 2000]{latter00} (NGC 7027)).  
The cryogen ran out
prematurely early in 1999 rendering the instrument no longer functional.
In early 2002, shortly after the last IAU PN Symposium, a cryocooler was installed that
brought the NICMOS instrument back to operation.   Since then, several new
projects have been carried out, as well as studies of data in the archive that
were obtained during the original cryogen lifetime. 

One example of new NICMOS imaging that was performed on a young PN is \cite{sahai05},
who imaged the OH/IR star IRAS 19024+0044 with NICMOS (1.1 and 1.6 $\mu$m) and
the ACS (at 0.6 and 0.8 $\mu$m). They found a compact multipolar PN with at least
six elongated, limb-brightened lobes originating from the central star.  The central
region is brighter at the longer wavelengths, with the peak in the 1.6 $\mu$m image
positioned near the center of symmetry of the lobes.  The NICMOS 
images reveal the region
near the core where there are dark lanes in the optical images.

Another result from new NICMOS imaging is the \cite{meixner05} study
of several regions in NGC 7293 (the Helix).
While the ACS was imaging the central region and main ring,
NICMOS was used in parallel to image positions on the main ring and
the halo.  In this nearby PN, they resolved the H$_2$ emission in 
the ring into individual molecular knots that have the appearance of small arcs with
their apex pointing toward the central star.  The arcs are similar to the cometary
knots previously observed in the central hole, but without the long tails.
At increasing radius, they found the H$_2$ emission becomes less structured. From
the H$_2$ images, they estimate a total neutral gas mass of the PN.

\section{IR Surveys (2MASS, MSX)}
Infrared imaging surveys such as 2MASS and MSX have become 
available since the last IAU PNe symposium and have provided a resource that has begun to 
be exploited for the study of PNe.  For example, Phillips \& Ramos Larios have in a 
series of papers (\cite{phillips05}; \cite{larios05}; \cite{phillips06}) used
the 2MASS database to study the near-IR properties of the PNe in the extended source
catalog, and to examine hot dust in these nebulae.  They report the presence of
hot dust halos for several of the PNe with large $K$ and $H$ band excesses, possibly due
to small grains.  
\cite{gauba03} used optical photometry along with 2MASS, MSX, and IRAS data to examine a 
sample of hot post-AGB candidates.  They modeled the SEDs and derived mass loss
rates, temperatures, distances, and sizes of the dust envelopes.

\section{Spitzer/IRAC and MIPS Imaging}

With the launch of the \textit{Spitzer Space Telescope}, an important 
new tool became available for the study of evolved stars and their ejecta.  The
three instruments on board \textit{Spitzer} -- IRAC (\cite[Fazio et al. 2004]{fazio04}), 
IRS (\cite[Houck et al. 2004]{houck04}) and MIPS (\cite[Rieke et al. 2004]{rieke04}) 
provide imaging capability from 3 -- 200 $\mu$m.  The IRS is a powerful 
spectrograph operating in the 5 -- 40 $\mu$m range,
and is reviewed in another paper at this symposium (\cite[Bernard-Salas 2006]{salas06}).  
In this review I focus on
\textit{Spitzer's} broad-band imaging capability at 3.6, 4.5, 5.8, 8, 24, 70, and 160
$\mu$m.  The instruments sample a spectral range that is difficult or impossible 
to observe from the ground, and can achieve $\mu$Jy sensitivities in a matter of 
minutes over large fields, making \textit{Spitzer}
the most sensitive platform ever for probing faint compact mid-IR  
sources and extended emission from PNe.  The relatively low 
spatial resolution ($\sim$2" 
for IRAC) means that \textit{Spitzer} is complementary 
to ground-based instruments that have been used to resolve bright, compact structures. 

\subsection{\textit{Spitzer} Survey Data}

There are several large surveys already in the \textit{Spitzer} public
archive that can be used for the study of PNe.  One of the largest is the 
GLIMPSE project (\cite[Churchwell et al. 2004]{church04}) 
which surveyed a large part of the galactic
plane with IRAC.  The MIPSGAL project (\cite[Carey et al. 2005]{carey05})
will map a similar area with the MIPS 24 and 70 $\mu$m bands.  The properties of known
PNe can be determined, and also searches for new nebulae performed
(e.g., \cite[Cohen et al. 2005]{cohen05}).  Another potentially useful data set is
the SAGE survey of the LMC (\cite[Meixner et al. 2006]{meix06}), which imaged a 7x7 
degree region with MIPS and IRAC.  Source catalogs and mosaics for SAGE 
are expected to be released in late 2006.  All data can be accessed via the 
\textit{Spitzer} web site at http://ssc.spitzer.caltech.edu.

\begin{figure}
\includegraphics[scale=0.35,angle=0]{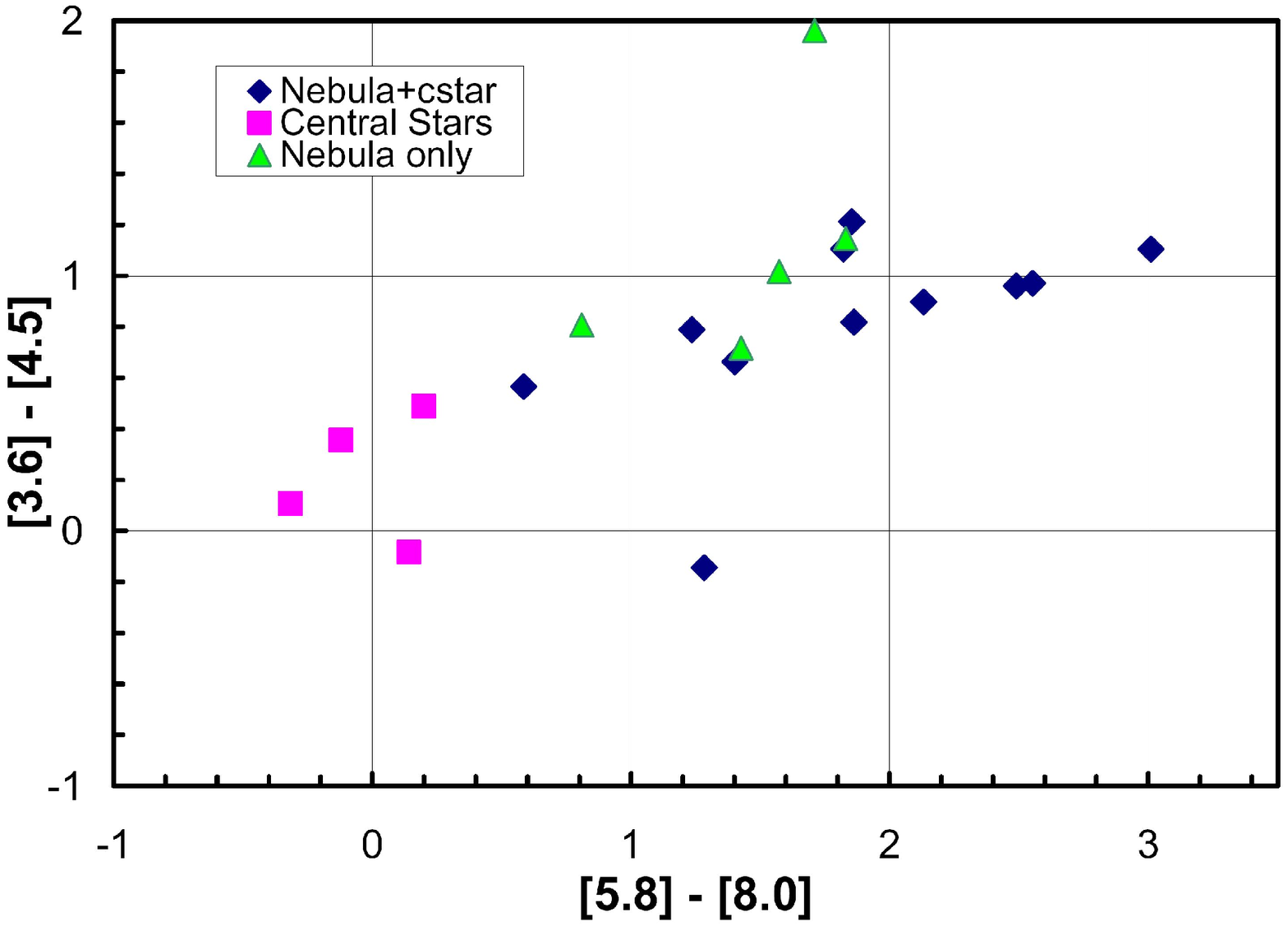}
\includegraphics[scale=0.34,angle=0]{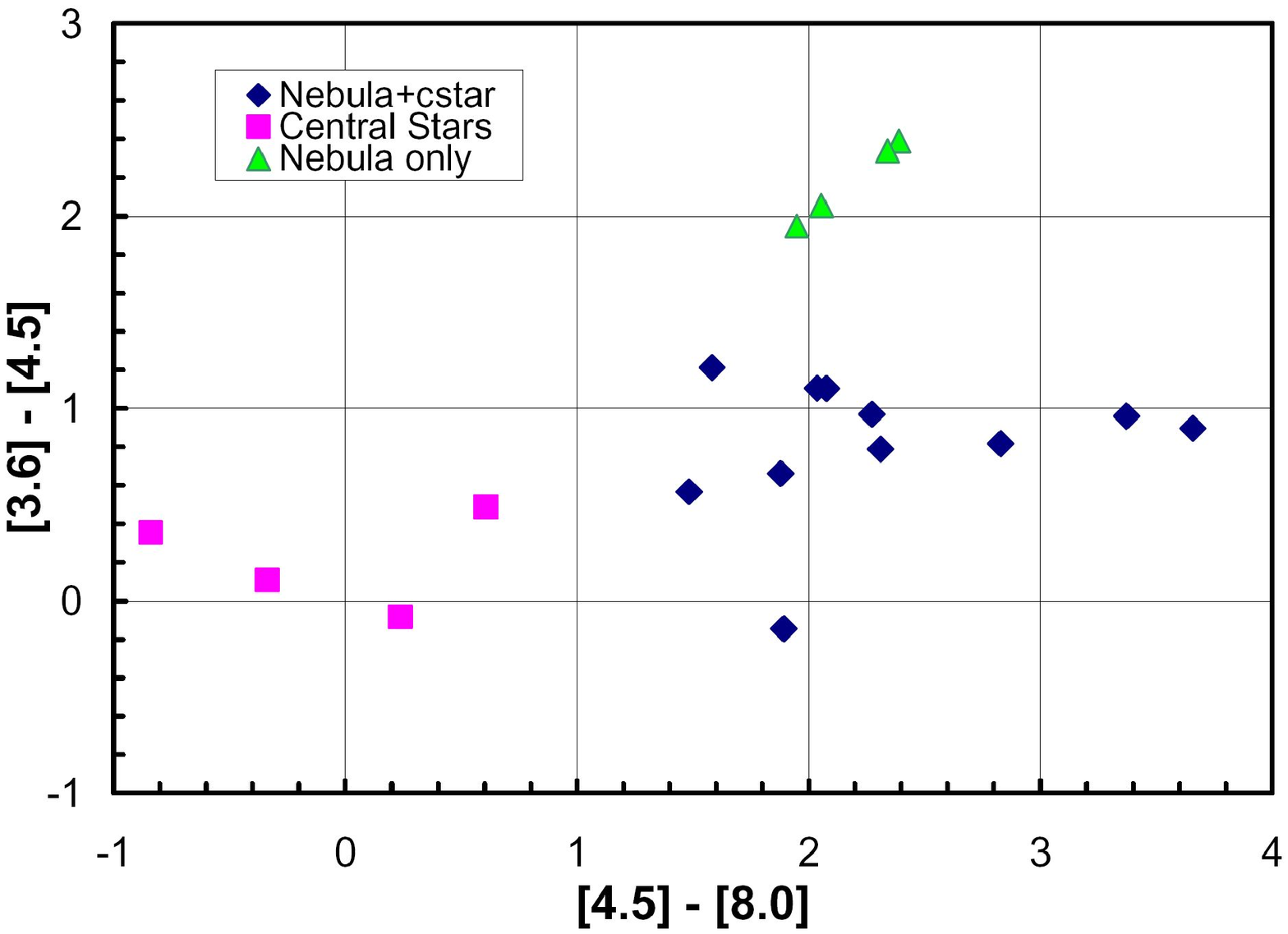}
\caption{IRAC color-color plots for some of the PNe in the GTO
sample.  The left plot shows the 4-band color-color plot, the plot
on the right shows the plot with bands 1, 2, and 4.  
The nebulae are red (brighter in bands 3 and 4) compared to main 
sequence stars and the PN central stars.  This is likely due to emission
in bands 3 and 4 from H$_2$ or forbidden line emission from the ionized
gas.  In some cases where the central star was detected, the star and
nebula were plotted separately.}\label{fig:colors}
\end{figure}

\subsection{IRAC survey of PNe}

The IRAC GTO program includes a project to image 35 PNe with IRAC.  A parallel
program was carried out in MIPS GTO time by W. Latter.  Some early
results have been presented (\cite[Hora et al. 2004]{hora04}; 
\cite[Hora et al. 2005]{hora05}; \cite[Hora et al. 2006]{hora06}), 
a sample are shown here.  Figure \ref{fig:colors} 
shows the IRAC colors for a number of PNe observed.  The nebular colors are in general 
red, especially comparing the 3.6 or 4.5 $\mu$m bands to the 5.8 and 8 $\mu$m.

\begin{figure}
\includegraphics[scale=0.67]{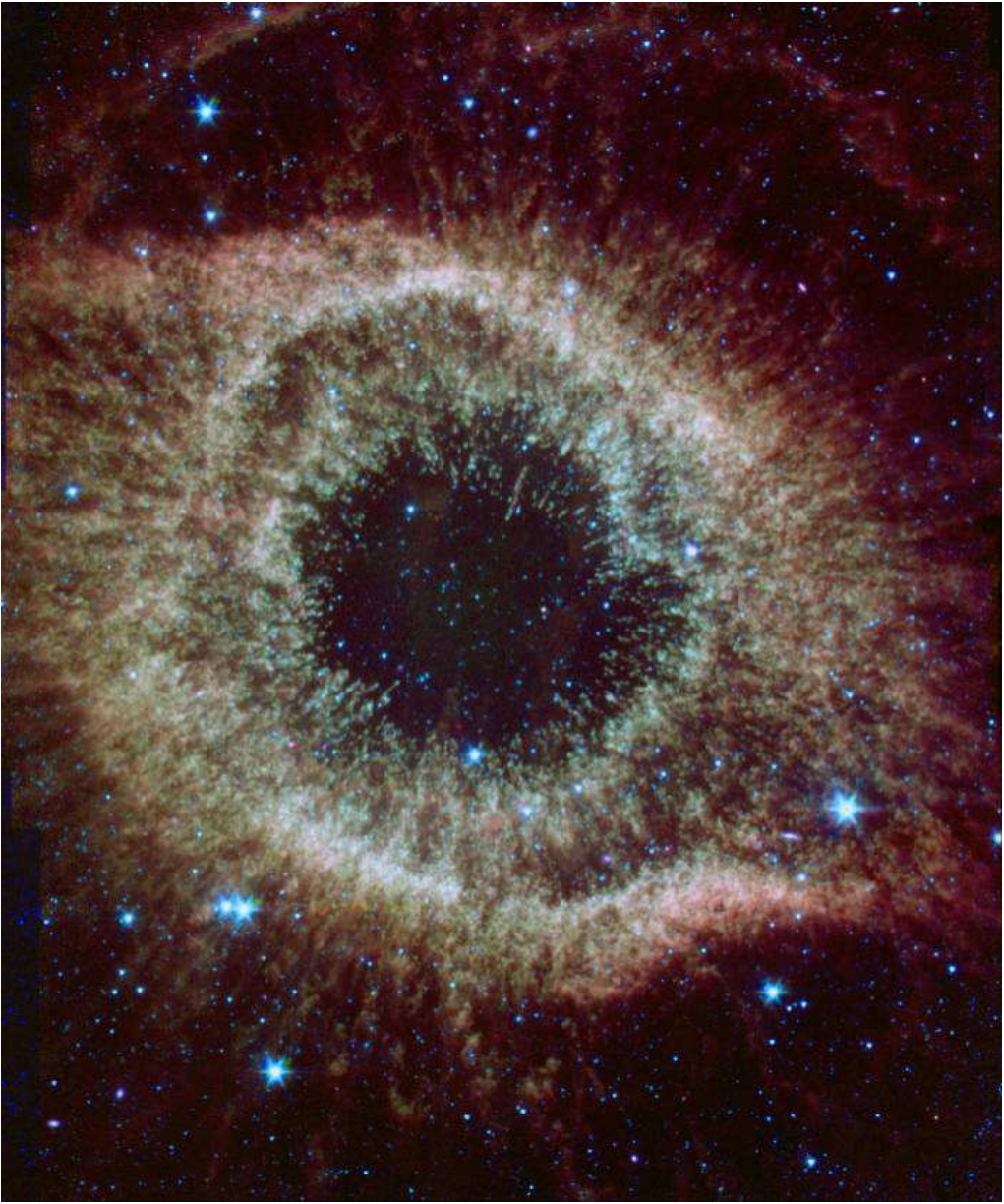}
\caption{IRAC image of the Helix, 
with the 3.6, 4.5, 5.8, and 8.0 $\mu$m 
bands are mapped to blue, green, orange, and red, respectively.  The image is approximately 24 $\times$ 26 arcmin in size.  The cometary knots are visible
inside the main ring; the knot tips are relatively brighter in the 3.6 and 4.5 $\mu$m 
bands, and the tails are brighter at 8 $\mu$m.  The nebular emission 
is dominated by H$_2$ in the IRAC bands. 
}\label{fig:helix}
\end{figure}

\begin{figure}
\includegraphics[scale=0.35]{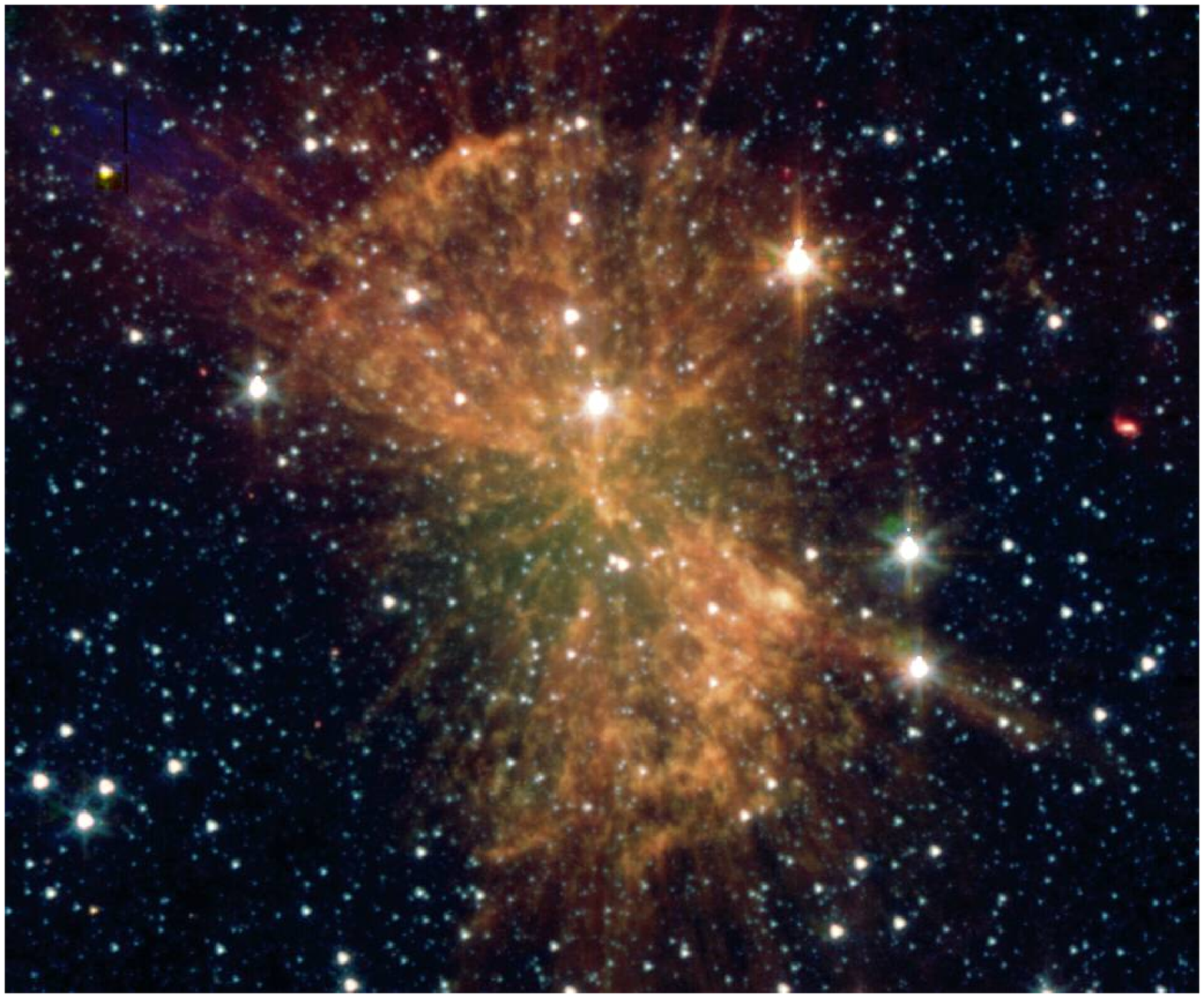}
\includegraphics[scale=0.32]{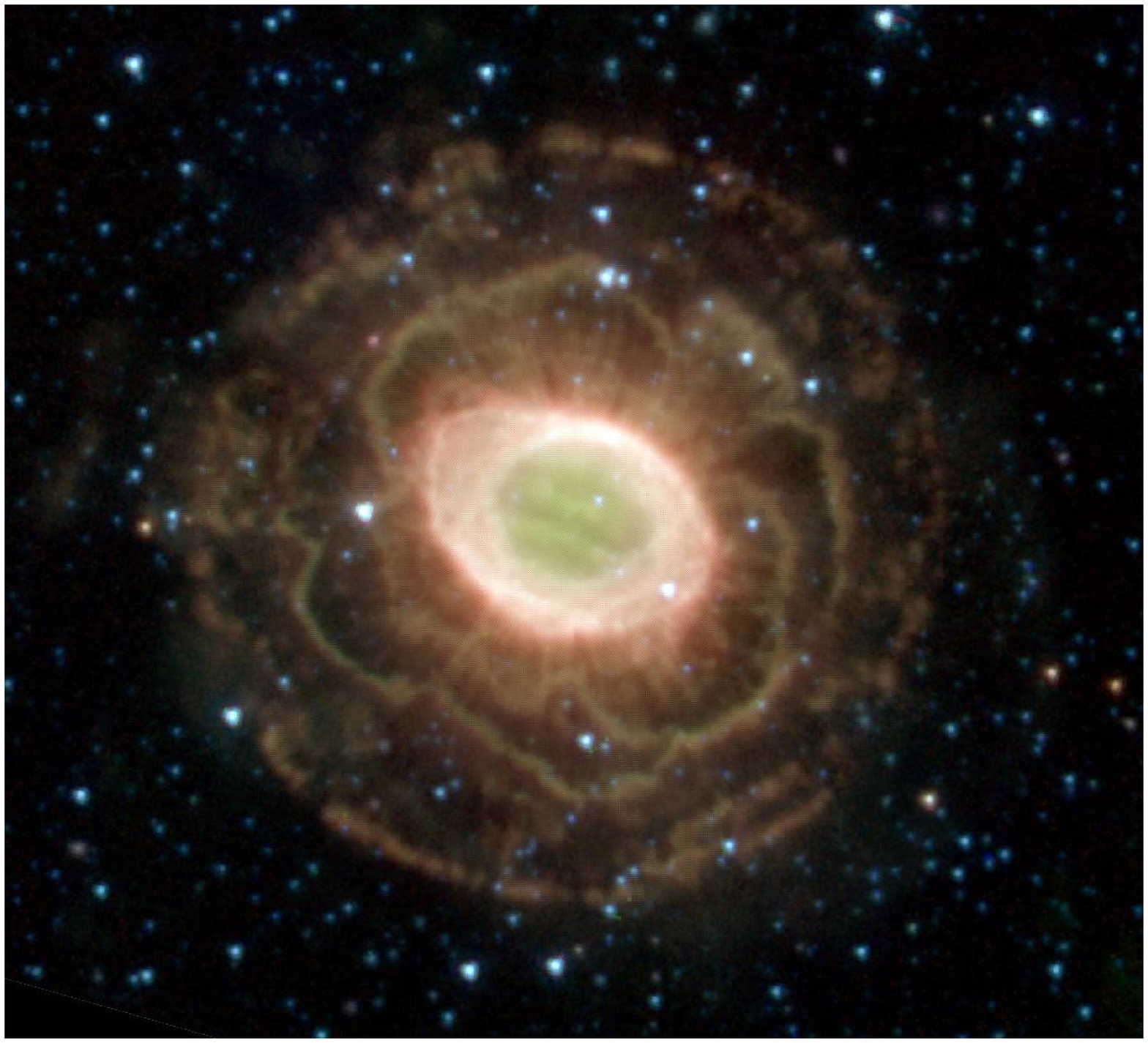}
\caption{ IRAC color images of PNe.  The 3.6, 4.5, 5.8, and 8.0 $\mu$m 
bands are blue, green, orange, and red, respectively.  Left: the Dumbbell (NGC 6853).
Right: the Ring (NGC 6720).  
}\label{fig:db_ring}
\end{figure}

The IRAC image of the Helix is shown in Figure \ref{fig:helix} (\cite[Hora et al. 2006]{hora06}).  
The four bands are mapped into colors of blue to red in order of wavelength.  The 
IRAC images and IRS spectra, along with 2.12 $\mu$m images, confirm the results of 
\cite{cox98} who mapped the Helix with \textit{ISO} and found that the mid-IR 
emission was dominated by lines of H$_2$.  Along with Cox et al. and \cite{meixner05}
we find that the H$_2$ line ratios are inconsistent with PDR models.  We find that they
can be fit by models of shock-excited H$_2$, with a small PDR component.  The IRAC 
emission is clumpy in the rings, each clump appearing as a small arc pointing back 
toward the central star, as in the 2.12 $\mu$m H$_2$ images.  
Outside of the ring, there are radial rays extending into
the halo.

Figure \ref{fig:db_ring} shows the IRAC images of the Dumbbell (NGC 6853)
and the Ring nebula (NGC 6720).
For the Dumbbell nebula, the appearance in the 
IRAC image is drastically different than in the optical. 
The IRAC emission is primarily from the roughly N-S 
dense equatorial region, and is 
dominated by the clumpy emission and radial rays that are strongest in the 
long wavelength channels.  The appearance at 8 $\mu$m is almost identical to
the 2.12 $\mu$m H$_2$ image, indicating a common origin.
In the Ring nebula, there are similarities to the
optical appearance, but the H$_2$ emission in the IRAC bands
highlights different parts of the
nebular structure (see also \cite[Speck et al. 2003]{speck03}).  
In the IRAC color image in Figure \ref{fig:db_ring},
the central region is slightly green from 
emission in the 4.5 $\mu$m bandpass, possibly due to atomic lines such as Br$\alpha$, 
[Mg IV] and [Ar VI] in that band.  The edge of the ring and the outer halo are 
orange-red, consistent with emission from H$_2$ in the 5.8 and 8 $\mu$m bands. 

An early MIPS imaging result was reported by \cite{su04}, who imaged NGC 2346 at 24, 70,
and 160 $\mu$m.  The distribution of the 24 $\mu$m emission was found to be very similar to the optical H$\alpha$ emission, except for a hot dust component that peaks
at the position of the central star.  This dust is likely responsible for the 
previously observed deep
fadings of the central star.  The 70 $\mu$m image shows a double-peaked morphology
in the narrow waistband of the nebula, indicating a possible dust torus structure. 
The 160 $\mu$m image shows that the nebula is contained in an extended cold dust envelope.

\section{Conclusions}\label{sec:concl}
Infrared imaging is an important probe of the molecular and dust components of 
PNe, as well as the ionized gas in regions that are obscured at shorter
wavelengths.  
Much of the recent work has focused on high spatial resolution studies of young
PNe, exploring the early development of nebula formation and searching for evidence
of the processes responsible for the nebula morphology.  New surveys and instruments
such as those on board \textit{Spitzer} have recently become available and are being
utilized to expand our knowledge of the formation and evolution of PNe.

\begin{acknowledgments}

This work is based in part on observations made with the Spitzer Space Telescope, 
which is operated by the Jet Propulsion Laboratory, California Institute 
of Technology under NASA contract 1407. Support for the IRAC instrument 
was provided by NASA under contract number 960541 issued by JPL.
\end{acknowledgments}

\begin{discussion}
\end{discussion}

\end{document}